\documentclass[aps,prl,twocolumn,superscriptaddress,longbibliography]{revtex4-1}
\usepackage{natbib}
\usepackage{graphicx}
\usepackage{dcolumn}
\usepackage{bm}
\usepackage{color}
\usepackage[usenames,dvipsnames,svgnames,table]{xcolor}

\begin{document}


\title{Hall, Seebeck, and Nernst Coefficients of Underdoped HgBa$_2$CuO$_{4+\delta}$ : \\
Fermi-Surface Reconstruction in an Archetypal Cuprate Superconductor}



\author{Nicolas~Doiron-Leyraud}
\email[]{ndl@physique.usherbrooke.ca}
\affiliation{D\'epartement de physique \& RQMP, Universit\'e de Sherbrooke, Sherbrooke, Qu\'ebec, Canada J1K 2R1}

\author{S.~Lepault} 
\affiliation{Laboratoire National des Champs Magn\'etiques Intenses, UPR 3228, (CNRS-INSA-UJF-UPS), Toulouse 31400, France}

\author{O.~Cyr-Choini\`ere}
\affiliation{D\'epartement de physique \& RQMP, Universit\'e de Sherbrooke, Sherbrooke, Qu\'ebec, Canada J1K 2R1}

\author{B.~Vignolle} 
\affiliation{Laboratoire National des Champs Magn\'etiques Intenses, UPR 3228, (CNRS-INSA-UJF-UPS), Toulouse 31400, France}

\author{G.~Grissonnanche}
\affiliation{D\'epartement de physique \& RQMP, Universit\'e de Sherbrooke, Sherbrooke, Qu\'ebec, Canada J1K 2R1}

\author{F.~Lalibert\'e}
\affiliation{D\'epartement de physique \& RQMP, Universit\'e de Sherbrooke, Sherbrooke, Qu\'ebec, Canada J1K 2R1}

\author{J.~Chang}
\affiliation{D\'epartement de physique \& RQMP, Universit\'e de Sherbrooke, Sherbrooke, Qu\'ebec, Canada J1K 2R1}

\author{N.~Bari\v si\'c} 
\affiliation{School of Physics and Astronomy, University of Minnesota, Minneapolis, Minnesota 55455, USA}

\author{M.~K.~Chan} 
\affiliation{School of Physics and Astronomy, University of Minnesota, Minneapolis, Minnesota 55455, USA}

\author{L.~Ji} 
\affiliation{School of Physics and Astronomy, University of Minnesota, Minneapolis, Minnesota 55455, USA}

\author{X.~Zhao}
\affiliation{School of Physics and Astronomy, University of Minnesota, Minneapolis, Minnesota 55455, USA}
\affiliation{State Key Lab of Inorganic Synthesis and Preparative Chemistry, College of Chemistry, Jilin University, Changchun 130012, China}

\author{Y.~Li}
\affiliation{International Center for Quantum Materials, School of Physics, Peking University, Beijing 100871, China}

\author{M.~Greven} 
\affiliation{School of Physics and Astronomy, University of Minnesota, Minneapolis, Minnesota 55455, USA}

\author{C.~Proust} 
\affiliation{Laboratoire National des Champs Magn\'etiques Intenses, UPR 3228, (CNRS-INSA-UJF-UPS), Toulouse 31400, France}
\affiliation{Canadian Institute for Advanced Research, Toronto, Ontario, Canada M5G 1Z8}

\author{Louis~Taillefer}
\email[]{louis.taillefer@usherbrooke.ca}
\affiliation{D\'epartement de physique \& RQMP, Universit\'e de Sherbrooke, Sherbrooke, Qu\'ebec, Canada J1K 2R1}
\affiliation{Canadian Institute for Advanced Research, Toronto, Ontario, Canada M5G 1Z8}
\date{\today}

\begin{abstract}

Charge density-wave order has been observed in cuprate superconductors whose crystal structure breaks the square symmetry of the CuO$_2$ planes, such as orthorhombic YBa$_2$Cu$_3$O$_{y}$ (YBCO),
but not so far in cuprates that preserve that symmetry, such as tetragonal HgBa$_2$CuO$_{4+\delta}$ (Hg1201).
We have measured the Hall ($R_{\rm H}$), Seebeck ($S$), and Nernst ($\nu$) coefficients of underdoped Hg1201 in magnetic fields large enough to suppress superconductivity. The high-field $R_{\rm H}(T)$ and $S(T)$ are found to drop with decreasing temperature and become negative, as also observed in YBCO at comparable doping.
In YBCO, the negative $R_{\rm H}$ and $S$ are signatures of a small electron pocket caused by Fermi-surface reconstruction, attributed to charge density-wave modulations observed in the same range of doping and temperature. We deduce that a similar Fermi-surface reconstruction takes place in Hg1201, evidence that density-wave order exists in this material.
A striking similarity is also found in the normal-state Nernst coefficient $\nu(T)$, further supporting this interpretation.
Given the model nature of Hg1201, Fermi-surface reconstruction appears to be common to all hole-doped cuprates, suggesting that density-wave order is a fundamental property of these materials.

\end{abstract}


\maketitle




There is a growing body of evidence that competing ordered states shape the phase diagram of cuprates and the identification of those states is currently a central challenge of high-temperature superconductivity.
In the La$_2$CuO$_4$-based cuprates, whose maximal $T_c$ does not exceed 40~K, the existence of unidirectional density-wave order involving spin and charge modulations, known as stripe order~\cite{Kivelson2003,Vojta2009}, is well established, as in La$_{1.6-x}$Nd$_{0.4}$Sr$_x$CuO$_4$ (Nd-LSCO)~\cite{Ichikawa2000} and La$_{1.8-x}$Eu$_{0.2}$Sr$_x$CuO$_4$ (Eu-LSCO)~\cite{Fink2011} for instance. This stripe order causes a reconstruction of the Fermi surface~\cite{Cyr-Choiniere2009,Daou2009,Chang2010,Laliberte2011}, and may be responsible for the low $T_c$.
The observation of a small electron pocket in the Fermi surface of underdoped YBCO~\cite{Doiron-Leyraud2007,LeBoeuf2007}, 
a material with a high maximal $T_c$ of 93~K, showed that its Fermi surface also undergoes a reconstruction at low temperature~\cite{Chakravarty2008,Taillefer2009}.
Comparative measurements of the Seebeck coefficient in YBCO and Eu-LSCO revealed a detailed similarity~\cite{Chang2010,Laliberte2011}, suggesting that Fermi-surface reconstruction (FSR) in YBCO is caused by some form of stripe order.

Charge density-wave modulations were recently detected in YBCO, via high-field nuclear magnetic resonance (NMR)~\cite{Wu2011} and X-ray scattering~\cite{Ghiringhelli2012,Chang2012,Achkar2012,Blackburn2012} measurements, in the range of temperature and doping where FSR occurs~\cite{LeBoeuf2011,Laliberte2011}. Although the detailed structure of these modulations remains to be clarified, there is little doubt that they are responsible for the FSR in YBCO.

The fundamental question, then, is whether such charge modulations are a generic property of cuprates. Because both the low-temperature tetragonal (LTT) structure of Eu-LSCO and the orthorhombic structure of YBCO distort the square CuO$_2$ planes and impose a preferred direction, charge modulations are perhaps triggered or stabilized by these particular forms of unidirectional distortion.
To answer that question we need to examine a cuprate material with square, undistorted CuO$_2$ planes. The model material for this is Hg1201, a tetragonal cuprate with the highest maximal $T_c$ of all single-layer cuprates (97~K)~\cite{Zhao2006, Barisic2008}, in which no charge or spin modulations have yet been reported.
In this Letter, we present measurements of the Hall, Seebeck, and Nernst coefficients in underdoped Hg1201 which reveal a FSR very similar to that seen in YBCO, demonstrating that some form of density-wave modulation is present in Hg1201.
This is strong evidence that charge density-wave modulations, as seen in YBCO, are a universal and fundamental property of underdoped cuprates. They compete with superconductivity and might also play a role in both the pairing mechanism and the anomalous scattering in these high-$T_c$ superconductors~\cite{Taillefer2010}.
%


\begin{figure}
\includegraphics[scale=2.3]{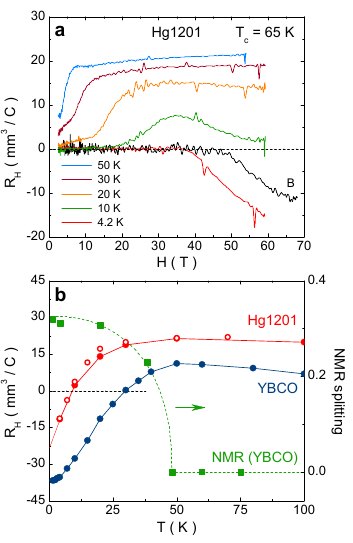}
\caption{
a)
Hall coefficient $R_{\rm H}$ of Hg1201 as a function of magnetic field $H$, in sample A at various temperatures, as indicated.
Also shown is an isotherm at $T = 4.2$~K measured on sample B (black curve, labelled B).
b)
Normal-state Hall coefficient $R_{\rm H}$ as a function of temperature, at $H = 53$~T (sample A; closed red circles, left axis)
and $H = 68$~T (sample B; open red circles, left axis).
Corresponding data are shown for YBCO at $p = 0.10$ and $H = 55$~T (blue circles, left axis; from~\cite{LeBoeuf2007}).
Note how in both materials $R_{\rm H}(T)$ drops at low temperature to become negative, a signature of 
Fermi-surface reconstruction~\cite{Taillefer2009}.
We reproduce the splitting of NMR lines in YBCO at $p = 0.10$ and $H = 28.5$~T, which
reveals the onset of charge order below $T_{\rm CO} \simeq 50$~K 
(green squares, right axis; from~\cite{Wu2011}).
}
%
\end{figure}



{\it Methods. --}
Two nominally identical high-purity single crystals of underdoped Hg1201 were measured (samples A and B), with $T_c \simeq 65$~K, prepared as described in refs.~\onlinecite{Zhao2006} and \onlinecite{Barisic2008}.
According to the $T_c(p)$ relationship for Hg1201 established in ref.~\onlinecite{Yamamoto2000}, our samples have a doping $p \simeq 0.075$.
We measured the Hall ($R_{\rm H} \equiv  \rho_{\rm xy} / H$), Seebeck ($S \equiv - V_x / \Delta T$), and Nernst ($\nu \equiv N / H \propto (V_y / \Delta T) / H$) 
coefficients, where $\rho_{\rm xy}$ is the transverse resistivity, and $V_x$ ($V_y$) is the longitudinal (transverse) voltage in the presence of a longitudinal temperature difference $\Delta T$.
For all measurements the magnetic field $H$ was applied perpendicular to the CuO$_2$ planes and the current (charge or heat) was within the plane.
Hall  measurements were performed in pulsed magnetic fields at the LNCMI in Toulouse up to $H = 68$~T, as described in ref.~\onlinecite{LeBoeuf2007}. 
The Seebeck coefficient was measured on sample A, as described in ref.~\onlinecite{Laliberte2011}, up to 28~T at the LNCMI in Grenoble 
and up to 45~T at the NHMFL in Tallahassee.
The Nernst coefficient was measured on sample A at Sherbrooke in a field of $H = 10$~T, as described in ref.~\onlinecite{Daou2010}.
%


\begin{figure}
\includegraphics[scale=2.3]{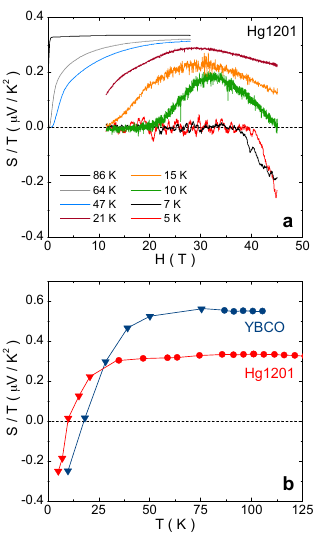}
\caption{
a)
Seebeck coefficient $S$ of Hg1201 plotted as $S/T$ vsÊ$H$, in sample A at various temperatures, as indicated.
b)
Normal-state Seebeck coefficient $S/T$ of Hg1201 as a function of temperature, at $H = 28$~T (red circles) and $H = 45$~T (red triangles).
Corresponding data are shown for YBCO at $p = 0.10$ in $H = 0$ (blue circles) and 28~T (blue triangles), from~\cite{Laliberte2011}).
}
%
\end{figure}



{\it Negative Hall and Seebeck coefficients. --}
In Fig.~1a, the Hall coefficient $R_{\rm H}$ is plotted as a function of magnetic field $H$ up to 68~T,
for different temperatures down to $T = 4.2$~K. 
All isotherms of sample A (B) saturate at high fields, beyond $H = 53$~T (68~T), except (including) at $T = 4.2$~K. 
In Fig.~1b, the high-field value of $R_{\rm H}$ is plotted versus temperature.
In Fig.~2a and 2b, we show the corresponding Seebeck data as a function of magnetic field and temperature, respectively.
As expected for a hole-doped material, both $R_{\rm H}$ and $S$ are positive at high temperature. However, with decreasing temperature they both start to fall below about $T \simeq 50$~K to eventually become negative below $T_0 = 10 \pm 1$~K.
This is our central finding: the low-temperature normal-state of Hg1201 has negative Hall and Seebeck coefficients.


\begin{figure}
\includegraphics[scale=2.4]{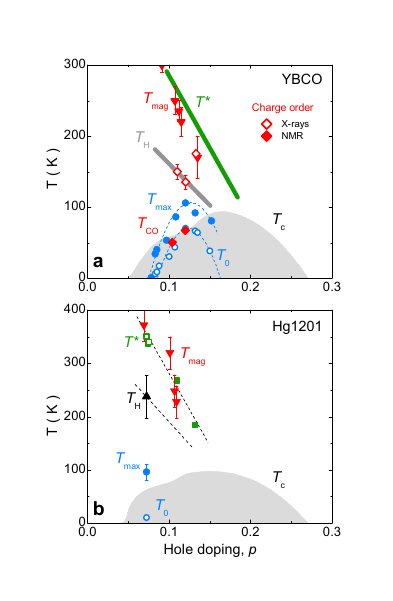}
\caption{
a)
Temperature-doping phase diagram of YBCO, showing the zero-field superconducting phase below $T_c$ (grey dome, from~\cite{Liang2006}) and 
the onset of $q=0$ magnetic order below $T_{\rm mag}$ detected by spin-polarized neutron scattering (down triangles, from~\cite{Fauque2006}).
Several characteristic temperatures of the transport properties are displayed:
the thick green and grey lines schematically represent the pseudogap temperature $T^\star$, 
from resistivity and Nernst data~\cite{Daou2010}, and $T_{\rm H}$, from $R_{\rm H}(T)$~\cite{LeBoeuf2011}, respectively.
$T_{\rm max}$ (full circles) and $T_0$ (empty circles) are determined from $R_{\rm H}(T)$~\cite{LeBoeuf2011} (see text).
We also show the onset of charge order at $T_{\rm CO}$ via NMR (full diamonds, from~\cite{Wu2011}) and the approximate onset of charge modulations via X-ray 
scattering (open diamonds, from~\cite{Ghiringhelli2012,Chang2012,Achkar2012}).
b)
Corresponding phase diagram for Hg1201, showing $T_c$ (grey dome, from~\cite{Yamamoto2000}), 
$T_{\rm mag}$ (down triangles, from~\cite{Li-Nature2008,Li-PRB2011}) and $T^\star$ from resistivity (full squares, from~\cite{Barisic2008,Li-Nature2008,Li-PRB2011};
open squares, from this work).
The characteristic temperatures $T_{\rm max}$ (full circles) and $T_0$ (empty circles) of $R_{\rm H}(T)$ are also shown.
All dashed lines in both panels are a guide to the eye.}
%
\end{figure}


In Fig.~1b and 2b, comparison with corresponding data in underdoped YBCO, with $T_c \simeq 57$~K ($p = 0.1$)~\cite{Laliberte2011,LeBoeuf2007}, reveals a striking similarity between the two cuprates.
In YBCO, there is compelling evidence that the negative $R_{\rm H}$ and $S$ at low temperature come from a small electron Fermi surface. This evidence includes quantum oscillations~\cite{Doiron-Leyraud2007,Jaudet2008} with a frequency $F$ and mass $m^\star$ which, at this doping, account precisely for the normal-state Seebeck coefficient $S$ at $T \to 0$, whereby $S/T \propto m^\star / F$, given that $S/T$ is negative~\cite{Chang2010,Laliberte2011}.
By analogy, we deduce that the Fermi surface of underdoped Hg1201 at low temperature also has an electron pocket.
This implies that it is reconstructed relative to its topology at high doping, where it is expected to be a single large hole-like cylinder, 
as observed in the single-layer tetragonal cuprate 
Tl$_2$Ba$_2$CuO$_{6 + \delta}$ at $p \simeq 0.25$~\cite{Mackenzie1996,Plate2005,Vignolle2008}.
In other words, the FSR in Hg1201 sets in below a critical doping $p^\star$ located somewhere between $p \simeq 0.08$ and $p \simeq 0.25$, as is the case for YBCO~\cite{LeBoeuf2011,Laliberte2011}, Eu-LSCO~\cite{Laliberte2011}, and Nd-LSCO~\cite{Daou2009,Cyr-Choiniere2009,Taillefer2009,Taillefer2010}. This quantum critical point at $p^\star$ in the normal-state phase diagram of Hg1201 marks the onset
of some density-wave order that breaks the translational symmetry of the lattice.

Our data also show that a transformation occurs upon cooling at fixed $p$, albeit smoothly, with no sign of a sharp transition.
The onset of FSR may be defined as the temperature $T_{\rm max}$ at which $R_{\rm H}(T)$ is maximal, although $R_{\rm H}(T)$ clearly starts to deviate downward from its high-temperature behavior well above $T_{\rm max}$, at a temperature labeled $T_{\rm H}$~\cite{LeBoeuf2011}. Our Hall data on Hg1201 sample B yield approximately 
$T_{\rm max} \simeq 100$~K and $T_{\rm H} \simeq 240$~K.
In YBCO, $T_{\rm max} \simeq 100$~K and $T_{\rm H} \simeq 120$~K at $p = 0.12$~\cite{LeBoeuf2011}.
In Fig.~3, $T_{\rm max}$, $T_{\rm H}$ and the Hall effect sign change temperature $T_0$ in YBCO and Hg1201 are plotted on their respective phase diagrams.

Given that the tetragonal structure of underdoped Hg1201 has no unidirectional character, our findings show that density-wave order is a generic tendency of the square CuO$_2$ plane, and therefore a phenomenon intrinsic to the physics of cuprates.
However, the precise nature of the density-wave order responsible for FSR in Hg1201 remains to be elucidated, {\it e.g.}, by X-ray scattering or NMR studies.

In YBCO at $p = 0.10$ and 0.12, a modulation of the charge density was detected in the CuO$_2$ planes by NMR measurements at high fields~\cite{Wu2011}. It was inferred to be unidirectional, with a period of $4a_0$, where $a_0$ is the lattice spacing, along the $a$-axis of the orthorhombic lattice for the ortho-II structure at $p=0.10$
(the pattern could not be determined for the ortho-VIII structure at $p=0.12$).
Note, however, that an additional modulation along the $b$-axis cannot be excluded.
In Fig.~1b, we reproduce the NMR data at $p = 0.10$ and see that the onset of charge order, at $T_{\rm CO} = 50 \pm 10$~K, coincides approximately with the downturn in $R_{\rm H}(T)$ for a similar doping ($p=0.10$). The same may be said of $S(T)/T$. Moreover, an increase in doping to $p = 0.12$ causes a parallel increase in both $T_{\rm CO}$~\cite{Wu2011} and $T_{\rm max}$~\cite{LeBoeuf2011} (see Fig.~3a). This is strong evidence that the FSR in YBCO is caused by this charge density-wave order~\cite{Doiron-Leyraud2012}.
In Eu-LSCO, unidirectional stripe-like charge order with a period $4a_0$ was detected by X-ray scattering~\cite{Fink2009}, with $T_{\rm CO} \simeq 40$~K at $p=0.10$ and $T_{\rm CO} \simeq 80$~K at $p=0.12$~\cite{Fink2011}, and linked to a drop in $R_{\rm H}(T)$~\cite{Cyr-Choiniere2009,Taillefer2009} and in $S(T)/T$~\cite{Chang2010,Laliberte2011}, again showing that charge order is causing the FSR~\cite{Doiron-Leyraud2012}.

Recent X-ray studies of YBCO in zero and low magnetic fields up to 17~T, however, have discovered incommensurate charge modulations (which may not be static) along both the $a$ and $b$ axes, with a period of $3.1~a_0$~\cite{Ghiringhelli2012,Chang2012,Achkar2012,Blackburn2012}.
As seen in Fig.~3, the onset of the X-ray scattering intensity appears to match $T_{\rm H}$, although these are gradual crossovers and there is no sharp anomaly in either the X-ray data or the transport data. Their relation to the charge order seen by NMR at high field below $T_{\rm CO}$ remains to be understood.
Recent high-field sound velocity measurements on YBCO at $p = 0.11$ detected the charge order below $T_{\rm CO}$, and showed that it must be a bi-directional charge-density-wave (and not domains of two uniaxial density waves)~\cite{LeBoeuf2012}. So the case of YBCO would appear to differ from the unidirectional charge-stripe scenario observed in the La$_2$CuO$_4$-based materials. 
But more work is needed to establish the differences and clarify whether these are fundamental.
It has been proposed that a bidirectional charge order is part of the explanation for the reconstructed Fermi surface of underdoped YBCO~\cite{Harrison2011}.
Given the striking similarity in the transport properties of YBCO and Hg1201, it is very likely that they host a  similar form of charge order.


\begin{figure}
\includegraphics[scale=2.4]{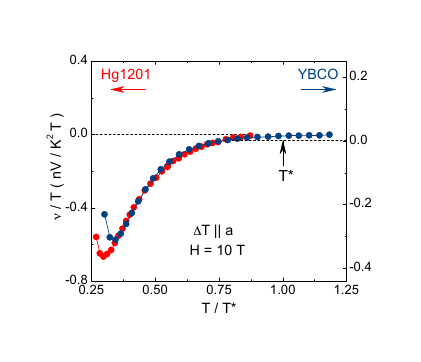}
\caption{
Nernst coefficient $\nu$ of Hg1201 ($T_c = 65$~K; red, sample A) and YBCO ($T_c = 66$~K, $p = 0.12$; blue, from~\cite{Daou2010}), plotted as $\nu/T$ vs $T / T^\star$, where $T^\star$ is the pseudogap temperature as defined in the main text. Here, $T^\star = 340$~K for Hg1201 and $T^\star = 250$~K for YBCO. The magnetic field $H =10$~T is along the $c$ axis and the heat gradient is along the $a$ axis. Note the different vertical scales for Hg1201 (red, left axis) and YBCO (blue, right axis). Below  $T / T^\star = 1.0$, the quasiparticle signal falls gradually to reach large negative values, in identical fashion in the two materials.}
%
\end{figure}


In YBCO, charge order competes with superconductivity~\cite{Ghiringhelli2012,Chang2012,Achkar2012}.
This is why $T_c$ falls when FSR sets in~\cite{LeBoeuf2011}, below $p \simeq 0.16$ (Fig.~3). The competition is strikingly manifest in recent measurements of the upper critical field $H_{c2}$ in YBCO~\cite{Grissonnanche2013}, which showed $H_{c2}(p)$ to have a local minimum where charge order exists.
Although less pronounced, a similar non-monotonic drop with underdoping is observed in the $T_c$ vs $p$ curve, for both YBCO~\cite{Liang2006} and Hg1201~\cite{Yamamoto2000} (see Fig.~3).
Certain features of the lattice structure may play a role in stabilizing the charge order, strengthening it more in some materials ({\it e.g.}  with the LTT structure).
This would impact on the competition between charge order and superconductivity, suppressing $T_c$ more effectively in Eu-LSCO (maximal $T_c \simeq 20$~K), where charge order exists at $H=0$, than in YBCO (where a magnetic field is needed to fully stabilize charge order~\cite{Wu2011,LeBoeuf2012}) or Hg1201 (maximal $T_c = 97$~K).
%


{\it Negative Nernst effect. --}
In underdoped YBCO, the Nernst coefficient $\nu(T)$~\cite{Chang2010,Daou2010} also provides hints of FSR. 
As seen in Fig.~4, the Nernst coefficient $\nu$ of YBCO at $p = 0.12$ is small and positive at high temperature, and it drops to large and negative values at low temperatures. (Note that unlike the Hall and Seebeck coefficients, the sign of the Nernst coefficient is not governed directly by the sign of the dominant charge carriers.)
This drop was shown to occur at the pseudogap temperature $T^\star$~\cite{Daou2010}, at which the in-plane resistivity $\rho_a(T)$ deviates from its linear temperature dependence at high temperature. Close to $T_c$, a positive signal due to superconductivity appears~\cite{Chang2012b}, but application of a large magnetic field suppresses this signal, revealing that the smooth drop in the normal-state $\nu/T$ continues monotonically down to $T=0$~\cite{Chang2010,Chang2011}.
The value of $| \nu/T |$ at $T \to 0$ is precisely that expected of the electron Fermi surface~\cite{Laliberte2011}, given its frequency, mass and mobility measured via quantum oscillations~\cite{Doiron-Leyraud2007,Jaudet2008}. 
In other words, the large negative Nernst coefficient in YBCO at low temperature is a consequence of FSR.

As shown in Fig.~4, the Nernst coefficient of Hg1201 is essentially identical to that of YBCO. 
When plotted versus $T / T^\star$, $\nu/T$ has the exact same temperature dependence in both materials. ($T^\star$ in Hg1201 is defined as in YBCO~\cite{Barisic2008,Li-Nature2008,Li-PRB2011}.)
We infer that the large negative $\nu(T)$ in Hg1201 is also a manifestation of FSR. (Note that in YBCO $\nu$ is anisotropic in the $ab$ plane~\cite{Daou2010,Chang2011}. 
In tetragonal Hg1201, where no such anisotropy is expected, the magnitude of $\nu$ lies between the $\nu_a$ and $\nu_b$ of YBCO.)


{\it Summary and outlook. --}
Our high-field measurements of Hall and Seebeck coefficients in the tetragonal single-layer cuprate Hg1201
reveal that its normal-state Fermi surface undergoes a reconstruction in the underdoped regime at low temperature,
which produces an electron pocket.
This is compelling evidence for the presence of a density-wave order that breaks translational symmetry.
The remarkable similarity of these transport properties with those of the orthorhombic bi-layer cuprate YBCO 
strongly suggest that the charge density-wave order observed in YBCO is also responsible for the FSR in Hg1201,
and is thus a generic property of hole-doped cuprates.
The presence of charge density-wave order in the midst of the phase diagram of cuprate superconductors raises some fundamental questions. Is the enigmatic pseudogap phase a high-temperature precursor of the charge order at low temperature? Is the quantum critical point for the onset of charge order responsible for the anomalous properties of the normal state, such as the linear-$T$ resistivity? Are fluctuations of the charge order involved in pairing?
Our findings in Hg1201 broaden the scope for exploring these questions by adding a clean archetypal cuprate to the list of materials that exhibit all the key properties of hole-doped cuprates, including superconductivity with a high $T_c$, a pseudogap phase with $q=0$ magnetic order~\cite{Li-Nature2008,Li-PRB2011,Li2012} (Fig.~3), and Fermi-surface reconstruction from charge density-wave order.


We thank the NHMFL for access to their 45~T hybrid magnet in Tallahassee.
The work at Sherbrooke was supported by a Canada Research Chair, CIFAR, NSERC, CFI, and FQRNT.
The work at the LNCMI was supported by PF7 EuroMagNET II and the ANR Superfield.
The work at the University of Minnesota (crystal growth, annealing, characterization and contacting of samples) 
was supported by the US Department of Energy, Office of Basic Energy Sciences.
J.~C. was supported by the Swiss National Science Foundation.


\bibliography{FSR_Hg1201}

\end{document}